\documentclass[10pt]{article}

\date{}

\usepackage{layouts}
\usepackage{balance}
\usepackage{url}
\usepackage{tabularx}
\usepackage{tabu}
\usepackage{dblfloatfix}
\usepackage{bm}
\usepackage{amssymb}
\usepackage{subfig}
\usepackage{graphicx}
\usepackage{mathtools}
\usepackage{authblk}

\usepackage{xcolor}

\definecolor{darkgreen}{HTML}{268827}
\newcommand{\green}[1]{{\color{black}#1}}

\usepackage{microtype}

\newcommand{\footnoteurl}[1]{\footnote{\url{#1}}}

\newcommand*{\fnref}[1]{\textsuperscript{\ref{#1}}}

\newcommand{\para}[1]{\smallskip\noindent\textbf{#1}}

\urldef\plosone\url{http://journals.plos.org/plosone/article?id=10.1371%2Fjournal.pone.0161636}

\title{Evidence of Online Performance Deterioration\\ in User Sessions on Reddit\thanks{Please cite the PlosOne Journal version of this article found at\newline \plosone.}}

\author[a,b,*]{Philipp Singer}
\author[c]{Emilio Ferrara} 
\author[c]{Farshad Kooti} 
\author[a,b]{Markus Strohmaier}
\author[c]{Kristina Lerman} 

\affil[a]{GESIS - Leibniz Institute for the Social Sciences}
\affil[b]{University of Koblenz}
\affil[c]{University of Southern California}
\affil[*]{philipp.singer@gesis.org}

\begin{document}
	
\maketitle

\begin{abstract}
	This article presents evidence of performance deterioration in online user sessions
	quantified 
	by studying a massive dataset containing over 55 million comments posted on Reddit in April 2015.
	After segmenting the sessions (i.e., periods of activity without a prolonged break) depending on their intensity (i.e., how many posts users produced during sessions), we observe a general decrease in the quality of comments produced by users over the course of sessions.
	We propose mixed-effects models that capture the impact of session intensity on comments, including their length, quality, and the responses they generate from the community. 
	Our findings suggest performance deterioration: Sessions of increasing intensity are associated with the production of shorter, progressively less complex comments, which receive declining quality scores (as rated by other users), and are less and less engaging (i.e., they attract fewer responses). 
	Our contribution evokes a connection between cognitive and attention dynamics and the usage of online social peer production platforms, specifically the effects of deterioration of user performance.
\end{abstract}

\section*{Introduction}
Performance deterioration following a period of sustained mental effort has been documented in settings that include student performance~\cite{sievertsen2016cognitive}, driving~\cite{Borghini2014measuring}, data entry~\cite{Healy04}, and exerting self-control~\cite{Muraven00}. Although the mechanisms for deteriorating performance are still debated~\cite{Boksem08,Kurzban13,Vohs08}, deterioration has been shown to be accompanied by physiological brain changes~\cite{Lorist05,Lim2010imaging,Pattyn2008psychophysiological}, suggesting a cognitive origin, whether due to mental fatigue, boredom, or strategic choices to limit attention. Outside of vigilance tasks, however, relatively little is known about whether and how this phenomenon affects online behavior. 
As our society becomes increasingly interconnected and people spend more time interacting through various online platforms, 
analyzing online performance is important for understanding how content is produced and consumed~\cite{ciampaglia2015production}, how information spreads~\cite{lerman2010information,yang2010predicting,wu2011says}, and how people decide what and who to pay attention to~\cite{weng2012competition,hodas2012visibility}.

\begin{figure}[t!]
	\centering
	\includegraphics[width=0.7\columnwidth]{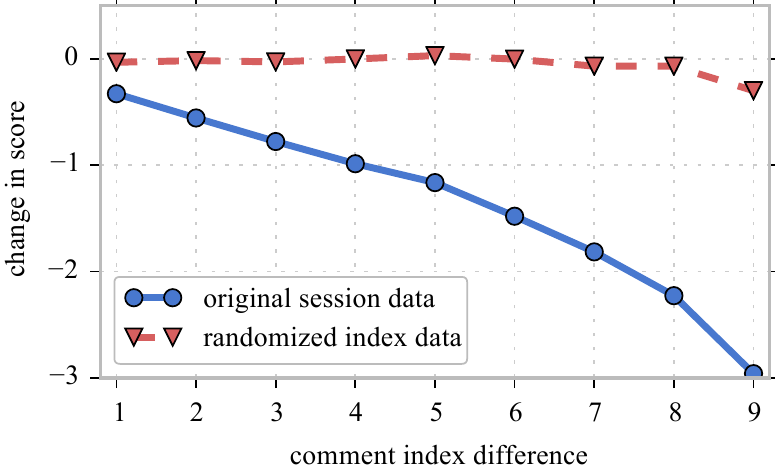}
	\caption{\emph{Performance of comments within sessions.} We show the average Reddit score for comments in sessions of length 10 (original session data, blue solid line). The average rating of each comment decreases starkly, by about 0.3 points for each comment after the first one in the session. This suggests the presence of (super linear) performance deterioration throughout user sessions. The effect disappears in randomized data having shuffled comments within sessions (red dashed line).}
	\label{fig:sessions}
	\vspace{-1em}
\end{figure}

\begin{figure}[t!]
	\centering
	\includegraphics[width=0.7\columnwidth]{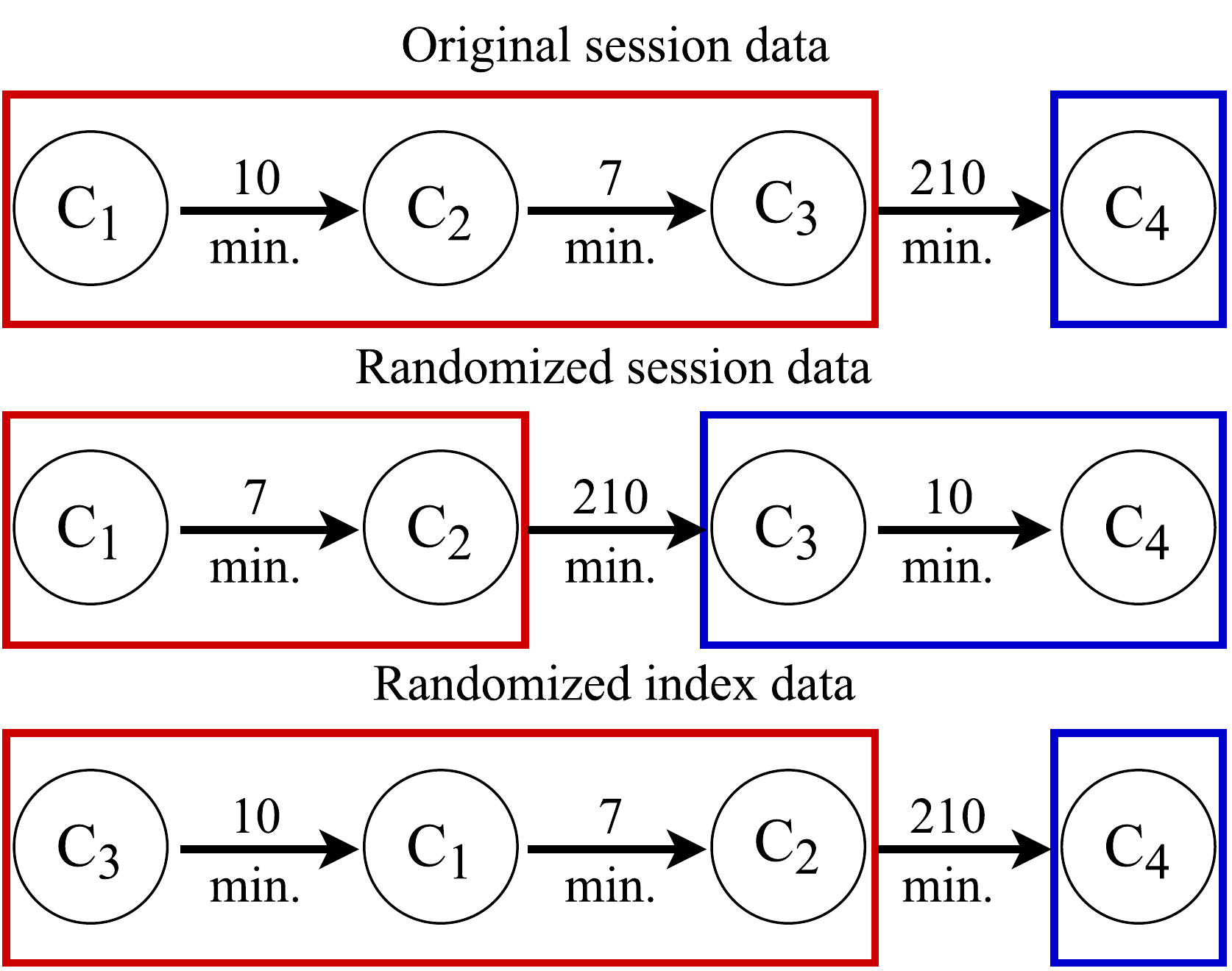}
	\caption{\emph{Sessions and randomization.} Circles represent comments $C_i$ and arrows depict the time difference $\Delta t_{i,j}$ between subsequent comments $C_i$ and $C_j$. Sessions are derived by breaking at time differences exceeding 60 min. Original data sessions are shown in the first row. The middle row shows  randomized sessions where time differences between comments are swapped for deriving new sessions while retaining the original order of comments. The bottom row depicts the randomized index data where sessions are retained but the order of comments within sessions is swapped.}
	\label{fig:randomization}
\end{figure}

In this work, situated under the broad umbrella of user behavior modeling \cite{allen1990user},  we study 
online performance on Reddit, a popular peer production and social news platform. 
We measure online peer production performance as the quality of comments produced by Reddit users over the course of a session, defined as a period of activity without a prolonged break.
The dataset we study contains over 55 million comments posted on Reddit in April 2015, and includes a variety of related meta-data, such as  time stamps, information about the users, and the score attributed by others to each comment.
We segment user activity into sessions, defined as periods of commenting  without a break longer than $60$ minutes, as suggested in \cite{halfaker2015user} (cf. Figure~\ref{fig:randomization}).
We link an individual's commenting performance over the course of a session to
different proxy measures for a comment's quality, such as its length, readability, the score it receives from others, and the number of responses it triggers.

Our analyses uncover deteriorating online performance over the course of user sessions, with a decline in quality of subsequent comments \emph{across} different proxy measures. 
Figure~\ref{fig:sessions} illustrates the decline in the average score received by comments posted during sessions with ten comments: the data shows that each subsequent comment receives a rating that is on average $0.3$ points lower than the preceding one. The size of this effect is quite large: It is equivalent to a 30\% probability increase of receiving a downvote to a comment, for each extra comment posted after the first one in the session.
\green{Additionally, we observe that users tend to start with higher quality comments the longer the sessions are.}
 To statistically study these effects, we design and implement mixed-effects models---allowing the incorporation of heterogeneous behavioral differences---which model the effect of session duration on the deterioration of online performance.

Our findings may be linked to effects of cognitive depletion: 
Exerting mental effort to compose a comment may diminish an individual's capacity to continue producing quality comments, whether through the loss of attention, mental fatigue, or simply the onset of boredom. 
Evidence also suggests that people, and other primates, have finite cognitive capacity for managing interpersonal relationships~\cite{Dunbar} limiting their amount of social interaction~\cite{Goncalves11,Miritello2013time}. Only recently, our research community started investigating the possible relationship between cognitive limits and online interactions, showing the impact of information overload on user behavior \cite{kooti2015evolution,Goncalves11,hodas2014simple,gomez2014quantifying}. Possibly, within-session deterioration of performance could explain the difficulty for users to continue exerting effort to discover information deeper in their social stream~\cite{hodas2012visibility,weng2012competition,ferrara2014online}. 
\green{Also, deterioration might be influenced by the passive content consumption within a session, e.g., replies by other users to own comments maybe being toxic or hateful leading to flame wars \cite{yasseri2012dynamics}.}
\green{The relation between the session length (i.e., number of comments) and the session's first comment's quality might also be explained by different starting capacities to make quality contributions, or, that the perceived quality of the first comment encourages users to produce more follow-up comments.}

Although unveiling the mechanism(s) behind  observed phenomena goes well beyond the scope of the current study, performance deterioration occurs throughout various critical daily activities, including learning (e.g., prolonged study sessions) and self-regulation (e.g., coping with stress, inhibition, refraining from behaving, or sticking to dietary restrictions). 
We believe that shedding light on the complex interplay between cognitive limits and individual performance can further our understanding of human behavior in many contexts.
Thus, showing initial evidence of online performance deterioration is important and
we expect this work to have implications for both computer and cognitive sciences communities.

\section*{Results}
\label{experiments}

Next, we present our findings on studying effects of session dynamics on online performance 
focusing on (\emph{i}) empirical observations, as well as by utilizing mixed-effects models on (\emph{ii}) performance at session start and (\emph{iii}) performance over the course of sessions.
We study, after pre-processing, around $40$ million Reddit comments posted in April 2015. We derive user sessions as periods of commenting activity without breaks longer than $60$ minutes as suggested in \cite{halfaker2015user} (cf. \emph{Materials and Methods} section and Figure~\ref{fig:randomization} for further details).
For measuring performance, we look at four proxies of comment quality:  text length, readability, the score a comment receives from others, and the number of responses it triggers. For comparison, we also study effects on two randomized session datasets as described in Figure~\ref{fig:randomization}.

\begin{figure*}[h!t!]
	\centering
	\subfloat{\includegraphics[width=\textwidth]{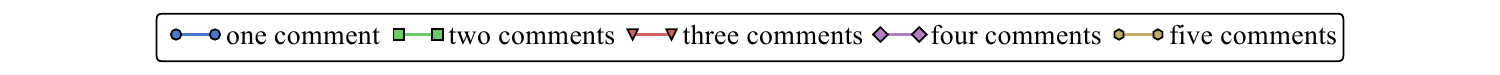}}
	\addtocounter{subfigure}{-1}
	
	\subfloat[Original session data]{\label{subfig:sessions_orig}\includegraphics[width=0.245\textwidth]{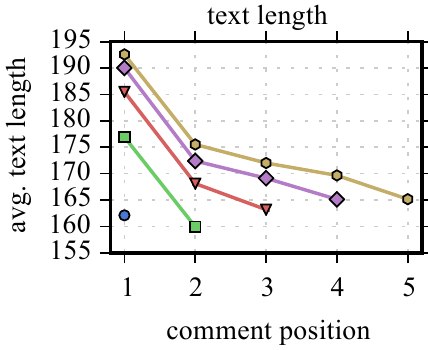}
		\includegraphics[width=0.245\textwidth]{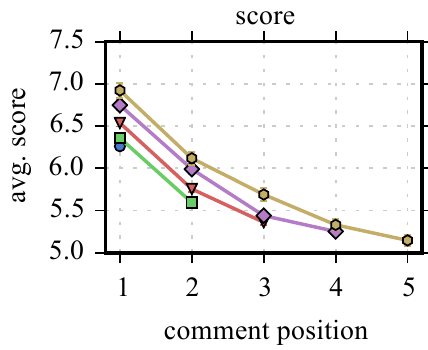}
		\includegraphics[width=0.245\textwidth]{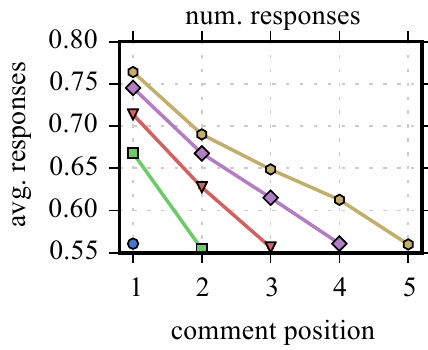}
		\includegraphics[width=0.245\textwidth]{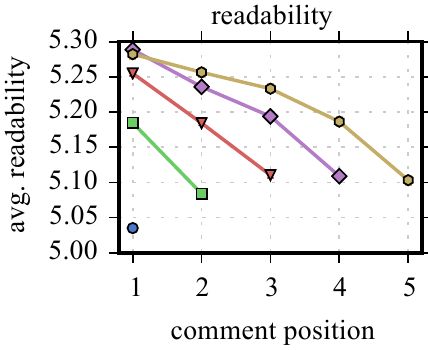}
	}\vspace{-1em}
	
	\subfloat[Randomized session data]{\label{subfig:sessions_random}\includegraphics[width=0.245\textwidth]{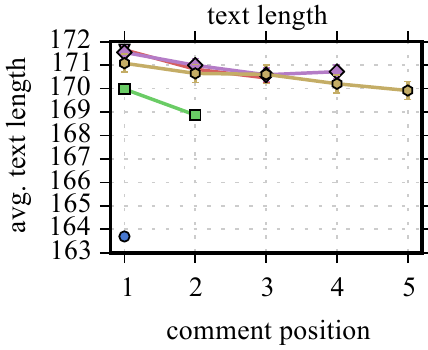}
		\includegraphics[width=0.245\textwidth]{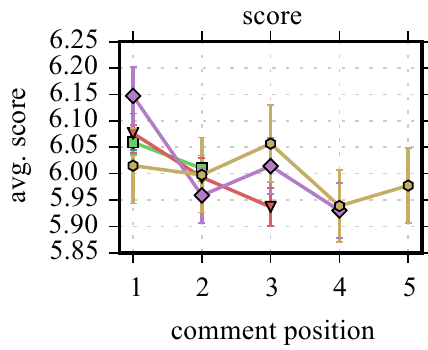}
		\includegraphics[width=0.245\textwidth]{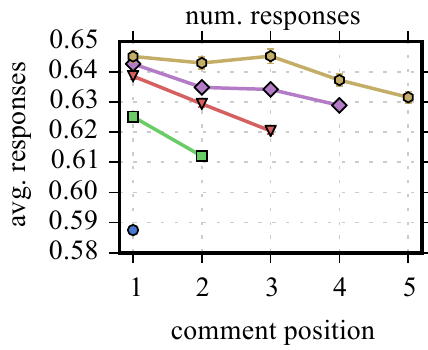}
		\includegraphics[width=0.245\textwidth]{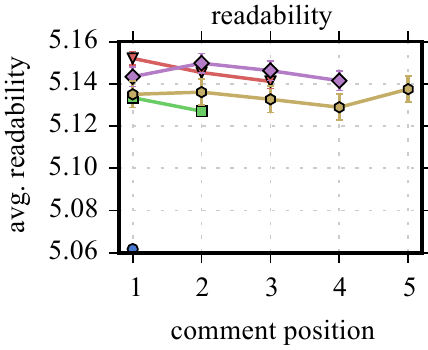}
	}\vspace{-1em}
	
	\subfloat[Randomized index data]{\label{subfig:sessions_random2}\includegraphics[width=0.245\textwidth]{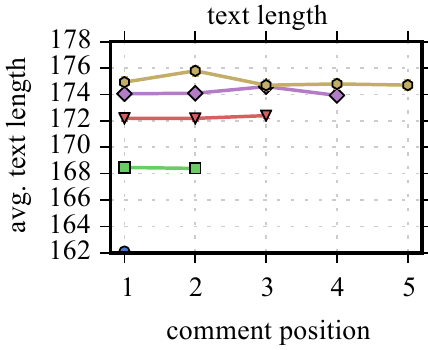}
		\includegraphics[width=0.245\textwidth]{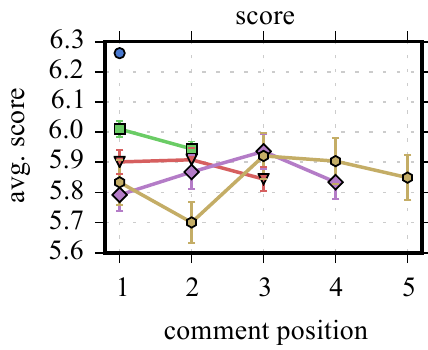}
		\includegraphics[width=0.245\textwidth]{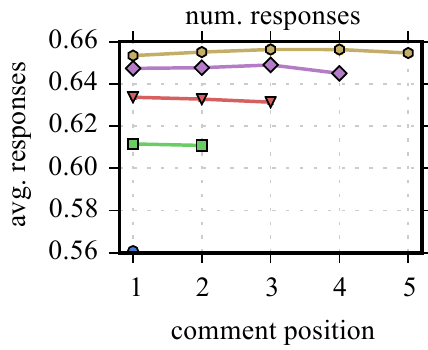}
		\includegraphics[width=0.245\textwidth]{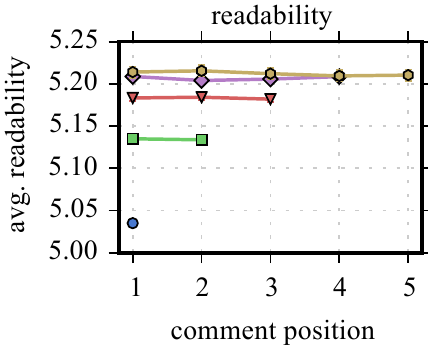}
	}
	
	\vspace{-0.5em}
	\caption{\emph{Empirical observations.} This figure visualizes the average of all four quality features of interest at their respective position in a session. The colors (different markers) indicate different session lengths (number of comments written in a session, $1$ up to a length of $5$). The x-axis depicts a comment's index within the session, and the y-axis gives the average feature value \green{(standard error bars of the mean are also depicted, but usually are very small and fall inside the data markers)}. The first row (a) depicts the original session data while the second (randomized session data) and third row (randomized index data) visualize results for the randomized data. The results indicate that earlier comments in a session tend to be of higher quality than later ones. Additionally, there appears to be a relation between the session length and the performance of the first comment in a session (stacking of lines). These clear patterns for the original data (a) mostly disappear for both of our randomized datasets (b,c). Overall, these empirical insights suggest performance deterioration over the course of sessions.}
	\label{fig:res}
	\vspace{-1em}
\end{figure*}

\subsection*{Empirical Observations}

Figure~\ref{fig:res} visualizes changes in online performance over the course of user sessions with respect to our quality features (comment text length, number of responses, score, and readability). Different colors and markers distinguish sessions of distinct length (i.e., number of comments written during the session) of up to a length of $5$.
The x-axis shows the session index of a comment, the y-axis shows the (population-wide) average of respective feature (with error bars). For example, in the first plot of Figure~\ref{subfig:sessions_orig}, the red triangle at  $x=2$ refers to the average text length of all comments written in second position of all sessions of length $3$.

Figure~\ref{subfig:sessions_orig} depicts the original session data of interest and suggests interesting dynamics in user behavior. First, all lines are stacked: The first comment of a longer session also starts out with a longer text, a higher score, more responses, and more complex text (evidenced by higher readability score). %
Second, all feature values decline throughout the course of a session hinting towards some form of performance deterioration. On average, the last comment of a session is shorter, receives a lower score and fewer responses, and is easier to read.

In contrast, these trends largely disappear in our randomized data---i.e., randomized session data shown in Figure~\ref{subfig:sessions_random} and randomized index data shown as in Figure~\ref{subfig:sessions_random2}. There is no clear decline in feature values of later comments 
in comparison to earlier comments in sessions. 
The reason why some lines (e.g., number of responses) in Figures~\ref{subfig:sessions_random}--\ref{subfig:sessions_random2} are still slightly stacked can be explained by our way of randomizing---see middle and bottom rows of Figure~\ref{fig:randomization}.
Some sessions (especially in the randomized session data) still stay partly, or sometimes even fully, intact, preserving the original session data. 
However, the effects are much reduced, for example, the average number of responses in the original data ranges between $0.55$ and $0.77$, while in the two randomized sets it ranges in the intervals [$0.58-0.65$] and [$0.56-0.66$] respectively.

Several considerations limit the conclusions we can draw from these empirical results.
First, the population-wide average feature value may not be fully indicative of user performance because  some distributions (length, score, responses) are heavy-tailed. Second, we have only visualized sessions up to a length of $5$. While visualizations including all lengths up to $10$ show similar trends (not shown here), more detailed analyses are necessary. 
Third, and most importantly, we have ignored the fact that our samples are not independent of each other as we repeatedly measure comments for individual redditors. Each user's behavior may be different, for example, one user may tend to write very long comments, while another one may prefer making shorter ones; mixing these different behavioral aspects in one analysis does not allow for specific inference about
performance deterioration.
We resolve some of these issues by using mixed-effects models incorporating individual differences (cf. \emph{Materials and Methods} section). We start with an (\emph{i}) analysis of the performance on the first comment in sessions, based on our observation of a potential stacking effect, and continue with (\emph{ii}) experiments on 
performance deterioration over the course of sessions
with respect to a potential decline in quality. 
\green{
In the main article, we only report the most appropriate models and the significant fixed effect coefficients. 
However, we make more extensive experimental results and model analytics available online\footnote{\label{notebooks}\url{https://github.com/psinger/reddit_depletion}} as well as in the supplementary material\footnote{The supplementary material of this article can be found at\newline \plosone.}. The provided R notebooks (S1-S8 Notebook) give insights about the experimental steps taken to come up with the appropriate models utilized in this work by examining a sample of 1 million data points. Additionally, we provide the complete regression output of the final models reported in this article in the supplementary material (S1-S8 Tables).
}

\subsection*{Performance at session start}

We hypothesized a relation between the length of sessions and their comments' respective quality; readily apparent in the stacking of lines in Figure~\ref{subfig:sessions_orig}.
We now statistically study this relation by focusing on the simplified question whether the length (number of comments) of a session has an effect on the performance of the very first comment in the session. 
We model the data with mixed-effects models specified as: $\text{feature} \sim 1 + \text{session length} + (1|\text{user})$.
The outcome (dependent) variable refers to one of our four quality features. The session length is the main fixed effect of interest. Additionally, we vary the intercept between users (random effect). 
For this analysis, we limit our data to only consider the very first comment in each session (around $23.5$M comments).
The detailed model analytics are openly available and can be found online\fnref{notebooks} and in the supplementary information
(S1-S4 Notebooks).

 The results (fixed session length effects) are summarized in Table~\ref{tab:res1} (cf. also S1-S4 Tables).
 As hypothesized in our empirical population-wide observations, the results indicate that there is a positive relation between the length of sessions (i.e., the number of comments) and their first comment's quality. This is imminent from resulting positive fixed effects coefficients meaning that an increase in session length leads, on average, to an increase of the first comment's text length, the number of responses it triggers, the score it receives and its Flesch-Kincaid grade level which corresponds to higher complexity of written text.

A potential explanation for the observed effect is that users start with different capacities to make quality contributions depending on how many more comments they plan to compose. Another (opposite) explanation could be that a higher performance of the first comment encourages users to produce more comments leading to longer sessions. While we believe the first explanation is more plausible---text length and readability are not based on external success measures, and responses  accumulate at a (somewhat) longer time scale---future studies should aim at answering these causal questions. 
Without resolving the nature of causality, the identified relation between session length and quality of the first comment has implications for the experiments we report below that model the dynamics of user performance during the sessions. We have now shown, empirically and statistically, a high heterogeneity between different sessions with respect to their length. Accounting for this (e.g., as a nuisance effect) in our models is thus necessary.  

\begin{table*}[h!] %
	\centering
	\caption{\emph{Mixed-effects model results.} In (a), the models study the effect of session length on the quality of the first comment $C_1$ in a session; i.e., data only contains the first session comments. In (b), the models investigate the effect of the session index $i$ on the quality of respective comment $C_i$; data includes all comments in sessions with more than a single comment.  Each table highlights the most appropriate models for each quality features based on extensive model analytics---lmer refers to linear mixed-effects models while glmer refers to generalized linear mixed-effects models. All coefficients are strongly significant as derived from model comparisons based on BIC statistics. }
	\subfloat[Performance at session start]{
		\label{tab:res1}
		\begin{tabularx}{\columnwidth}{|l|X|l|}
			\hline
			\emph{feature} & \emph{best model}  & \emph{coeff} \emph{(session length)} \\ \hline\hline
			text length & lmer (log-transform) & $+0.0342$ \\ \hline
			num. responses & glmer (Poisson, log link) & $+0.0685$\\ \hline
			score & glmer (Poisson, log link, constant added for positivity) &  $+0.00015$\\ \hline
			readability & lmer & $+0.0478$\\ \hline
		\end{tabularx}
	}%
	\\
	\subfloat[Performance over the course of sessions]{
		\label{tab:res2}
		\begin{tabularx}{\columnwidth}{|l|X|l|}
			\hline
			\emph{feature} & \emph{best model}  &\emph{coeff} \emph{(session index)} \\ \hline\hline
			text length & lmer (log-transform) & $-0.0205$ \\ \hline
			num. responses & glmer (Poisson, log link) & $-0.0640$\\ \hline
			score & glmer (Poisson, log link, constant added for positivity) &  $-0.00028$\\ \hline
			readability & lmer & $-0.0410$\\ \hline
		\end{tabularx}
	}
	\label{tab:res} 
	\vspace{-2em}
\end{table*}

\subsection*{Performance over the course of sessions}

We now turn our attention to the %
dynamics of user performance 
in sessions on Reddit. Our empirical insights so far have suggested a performance decline throughout the course of a session.
We statistically study this hypothesis by investigating whether the index of a comment (relative position in session) has an effect on the quality of the respective comment. 
 To that end, we apply mixed-effects models specified as: $\text{feature} \sim 1 + \text{session index} + \text{session length} + (1|\text{user})$.
Again, the dependent variable refers to one of our four quality features. The session index is the main fixed effect that we are interested in for studying performance declines. 
Our models include an additional nuisance effect controlling for individual session lengths as suggested by our previous experiments---model analytics confirm the importance of this factor. An additional random effect models the variations of the intercept between different authors.
For this analysis, we consider all data for sessions having more than a single comment (around $24.5$M comments).
The detailed model analytics can be found online\fnref{notebooks}  and in the supplementary information (S5-S8 Notebooks).

We summarize the main results in Table~\ref{tab:res2} and again focus on the fixed session index effect (cf. also S5-S8 Tables). The results now indicate a negative effect of the session index on our respective quality features indicated by the four negative coefficients. This means that with duration of a session, the quality of comments decreases on average. The next comment in a session is of shorter text length, triggers less responses and a fewer score, as well has a lower Flesch-Kincaid grade level indicating easier complexity of written text. This argues for performance deterioration throughout the course of user sessions on Reddit.

To further confirm observed effect, we  repeated the above experiments on the randomized data. For both the randomized session and index datasets, the session index effect is not significant for all features of interest indicating no performance depletion effect in the randomized data (cf. S5-S8 Notebooks).
This is in contrast to  real session data analyzed above and also confirms that the effects do not simply arise as a result of the \emph{order} in which comments are made, but their \emph{order within} a session.

\section*{Discussion}

Our work presents novel evidence of performance deterioration during prolonged online activity. %
By analyzing Reddit, a popular online social network that attracts millions of users, %
we showed that sessions with more activity are significantly associated with production of lower quality content, as measured by the length of the comment posted, its readability score, its average score and the number of responses it receives. In light of these findings, we developed a mixed-effects model that captures online performance deterioration. 
The code and results for all model analytics are available online\fnref{notebooks} and available in the supplementary information (S1-S8 Notebooks and S1-S8 Tables).

Our analysis can be expanded in several directions.
For example, we have only accounted for the basic differences between distinct Reddit users in the mixed-effects models. 
Yet, a much more nuanced analysis of 
heterogeneous effects of online performance deterioration would be warranted. 
 One interesting direction involves understanding whether all individuals exhibit the same levels of performance deterioration, or whether these effects vary from user to user. For example, we might find that all users consistently exhibit deterioration or that different subgroups of users exist, where some users might even show improvements in performance over time. Neuroscience studies found individual differences in working memory and other cognitive activities in the human brain~\cite{vogel2004neural}.
However, it remains unclear from a physiological standpoint whether capacity to process or produce information  varies from person to person~\cite{marois2005capacity}. 
Online performance deterioration may also depend on acquired experience (as a form of cognitive dexterity) with a system.
A new, and thus unfamiliar, user in a system may experience faster performance deterioration than an experienced user, because e.g., the cognitive or attention cost associated with the same operations may be experience-dependent (this is particularly true for information discovery and content production activities).
A computational study of online performance in this direction could be very valuable. 

Additionally, other hypotheses can be studied, such as that performance deterioration depends on the topic (politics vs. funny images), the time of the day, or the intensity of sessions (shorter average time differences between comments). 
A further aspect to consider is, that we have considered all comments posted to Reddit as equal, meaning that we did not distinguish between those comments posted at the root of a comment hierarchy and those posted further down the hierarchy.
Future research in that direction is necessary to better understand observed deterioration effect. For example, top-level comments might generally be of higher quality than low-level comments, or performance deterioration might be stronger for successive posts in the same submission thread compared to comments across submissions. \green{Also, the position of a comment in the hierarchy also influences its visibility to others which might have an impact on perceived quality.}
These and similar questions can be studied by our proposed models. They are highly adaptable and fixed and random effects can be utilized to model these potential heterogeneous effects; for example, including a random effect allowing the deteriorating effects to vary between users could already allow us to make further inference about individual differences.

\green{Furthermore, the set of quality features can be extended arbitrarily and also investigated more closely. In this work, we have focused on two features that are static (text length and readability) and two features that express the perception of the content by others (score and number of responses). Specifically the latter category of features warrants future studies, e.g., in light of potential social influence bias (herding) effects \cite{muchnik2013social}. Yet, also other categories of quality features might be of interest, such as the sentiment of the comment.}

Although our study was confined to Reddit, performance deterioration may generalize to other online activities. Future studies are needed to identify  the mechanisms leading to observed deterioration, whether through the loss of attention, mental fatigue, or simply the onset of boredom. Regardless of the causes, understanding  the complex  interplay between individual's cognitive limits and dynamic behavior is key to optimizing individual---and collective---performance in peer production and other online systems.

\section*{Materials \& Methods}
Here, we thoroughly describe utilized data, corresponding pre-processing steps, and statistical mixed-effects modeling approach.

\subsection*{Data}

For studying performance deterioration we utilized a publicly available dataset\footnote{\label{data}\url{https://www.reddit.com/r/datasets/comments/3bxlg7/i_have_every_publicly_available_reddit_comment}} containing all comments (nearly $1.7$ billion) ever written on Reddit starting from the first one on October 17, 2007 to the last one at the end of May 2015. For our experiments, we extracted a smaller sample that limits the data to all comments posted in April 2015. An advantage of this limited data is that we do not need to additionally account for changes in Reddit's platform not only in its interface, but also in its voting mechanisms as well as the general usage patterns of users on the site \cite{singer2014evolution}. Our results are robust to sample data from other months showing similar observations.

\subsection*{Quality features}
\label{features}

For measure online performance, we studied the following comment quality features.

\para{Text length.}
This feature counts the number of characters in a comment and is an indicator for its textual length. 
Each URL in a comment accounts for one additional character. 
The overall mean of text lengths is $\mu=168.08$, the median is $m=86.00$, and the standard deviation is $\sigma=281.88$.

\para{Score.}
The score is a measure the perception of other users and is 
the difference between their up- and downvotes (the starting score is $1$). All ratings can be summarized by the mean $\mu=6.05$, the median $m=1.00$ and the standard deviation  $\sigma=51.57$.

\para{Number of responses.}
We see the number of replies a comment triggers as a proxy for engagement and a comment's success. We only count direct replies in the comment hierarchy. 
The mean number of responses is $\mu=0.61$, the median is $m=0.00$ and the standard deviation is $\sigma=1.44$.

\para{Readability.}
George Klare provided the original definition of readability~\cite{klare1963measurement} as ``the ease of understanding or comprehension due to the style of writing''. For measuring readability of Reddit comments, we use the so-called \emph{Flesch-Kincaid grade level} \cite{kincaid1975derivation} representing the readability of a piece of text by the number of years of education needed to understand the text upon first reading; it contrasts the number of words, sentences and syllables.
It is defined as follows:

\begin{align*}
0.39\ \left(\frac{\text{total words}}{\text{total sentences}}\right)+11.8\ \left(\frac{\text{total syllables}}{\text{total words}}\right) - 15.59
\end{align*}

The lowest possible grade is $-3.4$ which e.g., emerges for comments that only contain a sentences having a single syllable such as ``OK'', only a single URL or only emoticons. We set the maximum Flesch-Kincaid grade to be $22$. 
Simply put, a higher Flesch-Kincaid grade indicates higher readability complexity of a given comment. 
The overall mean of the Flesch-Kincaid grade is $\mu=5.12$, the median is $m=4.91$ and the standard deviation is $\sigma=4.61$.

\para{Correlation of features.}
Most of the features are not strongly correlated (Pearson's $\rho$) with each other; however, we can identify two special cases. First, readability and text length have a correlation of $\rho=0.296$, which is not surprising given that  shorter texts are easier to read, which is accounted for in the Flesch-Kincaid grade level formula. 
Second, the two success features score and number of responses have a correlation of $\rho=0.558$, meaning that comments that get a high score also tend to receive more replies.
However, overall, these correlation results indicate that each feature represents interesting aspects on its own. 
\green{All correlation coefficients are strongly significant (p-values close to zero) for a significance test with the null hypothesis stating no correlation (also accounting for multiple comparison by e.g., Bonferroni adjustment).}

\begin{table}[h!] %
	\caption{\emph{Pearson correlation between features.}}
	\centering
	\begin{tabular}{|c|c|c|c|c|}
		\hline
		& \small{text length}  & \small{readability} & \small{responses}   &  score \\ \hline \hline
		\small{text length}      &      1.000    &        0.296   & 0.072& 0.005\\ \hline
		\small{readability}   &  0.296        &    1.000  & 0.043 & 0.005\\ \hline
		\small{responses}    &        0.072     &      0.043  &   1.000& 0.558\\ \hline
		score           &         0.005       &       0.005   &  0.558 & 1.000\\ \hline
	\end{tabular}
	\label{tab:corr} 
	\vspace{-1em}
\end{table}
\subsection*{Sessions}
\label{sessions}

\begin{figure}[!t]
\centering
\includegraphics[width=0.8\columnwidth]{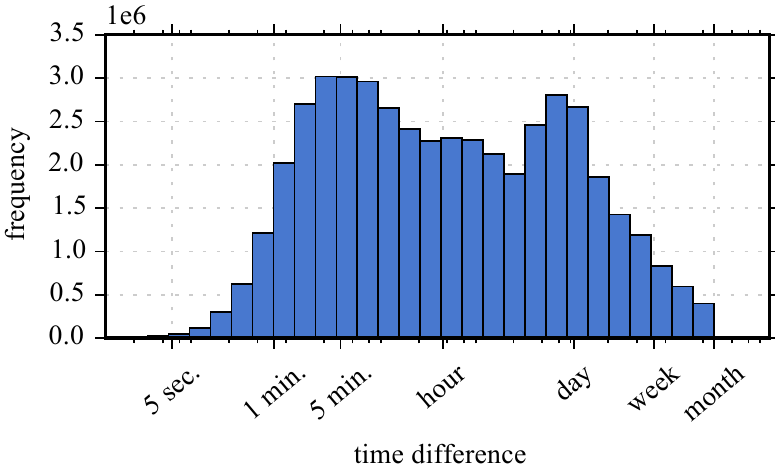}
\caption{\emph{Time differences between consecutive comments of users on Reddit.} \green{The x-axis depicts the time differences between consecutive comments (tick labels refer to major ticks) and the y-axis illustrates respective frequency.} The log-scaled histogram shows a peak for very short time scales (minutes) and very long ones (1 day) suggesting daily routines. A natural valley emerges between both peaks arguing for the choice of a one hour break between comments for sessions.}
\label{fig:time_difference}
\end{figure}

We decided to take the time differences between consecutive comments as session indicators. 
To that end, we followed the approach advised in \cite{halfaker2015user} where a strong regularity in how social media users initiate events across several different platforms was identified. Authors argue that a good rule-of-thumb is an inactivity threshold of $60$ minutes to separate sessions. However, as postulated, we first visually and analytically inspect the log-scaled histogram of time differences between consecutive comments (after cleaning comments, before filtering sessions) as depicted in Figure~\ref{fig:time_difference}. Similar to the results presented for other platforms \cite{halfaker2015user,geiger2013using}, there is a peak for very short time scales (minutes) and a peak for time differences of one day suggesting daily routine. 
By fitting a Gaussian Mixture Model (using EM-algorithm, log-normal mixture) with two components to the log-transformed data, we end up with the two means $\mu_1=6.85 \text{min.}$ and $\mu_2=794 \text{min.}$
A natural valley is visible between the two peaks and thus, combined with the results from the log-normal mixture fitting, we follow the rule-of-thumb of \cite{halfaker2015user} and pick a time difference $\Delta t_{i,j}$ of one hour between consecutive comments $C_i$ and $C_j$ to separate sessions.
Note that other (similar) choices  of break time (e.g., $30$ or $90$ minutes) produce similar inference. 

\subsection*{Data pre-processing}
We took several steps for pre-processing and cleaning the data. 
\green{First, we removed users from our data based on these rules:
	(\emph{i}) They have posted the exact same comment more than $100$ times, (\emph{ii}) their username is part of an unofficial Reddit bot list\footnote{\url{https://www.reddit.com/r/autowikibot/wiki/redditbots}}, or (\emph{iii}) their account has been deleted; this accounts for around $4.5$M comments.}
Second, we deleted all sessions containing at least one comment (\emph{i}) that has been deleted, (\emph{ii}) that is completely empty, or (\emph{iii}) that contains characters that are not in the ASCII character set (e.g., Chinese characters)---accounting for additional $3$M comments.
Finally, we removed all sessions containing more than $10$ comments accounting for around $7.25$M allowing
for easier experimental tractability and the removal of further bot accounts. Note though that the inclusion of these sessions into the experiments does not change the main observations of this paper. 
Our final dataset contains  $40,064,930$ comments produced by $2,669,969$ different users and posted in $47,462$ different subreddits.

\subsection*{Randomizing sessions}
For comparison, we created two randomized datasets to which we applied our analysis. The first baseline---which we call \emph{randomized session dataset}---attempts to preserve as much information as possible while randomizing the process of deriving user commenting behavior sessions. To do so, we  shuffled the time differences $\Delta t_{i,j}$ between consecutive comments made by each user, but preserved all other features, including the temporal order of comments. Then, we simply  derived  user activity sessions based on  shuffled times. An example is provided in Figure~\ref{fig:randomization} (middle row). This baseline dataset is very conservative in terms of randomization and retains many original sessions. For example, many parts of a session stay intact as only the short time differences are potentially swapped, which does not alter the sessions. The second baseline---which we call \emph{randomized index dataset}---keeps the sessions intact, but randomizes the order of comments inside each session (e.g., exchanging $C_1$ by $C_3$). Thus, it does not preserve the original order of comments; see Figure~\ref{fig:randomization} (bottom row). 
Multiple randomization iterations did not alter the results.

\subsection*{Mixed-effects models}
\label{multilevelmodels}

For statistically modeling performance deterioration, we utilized \emph{mixed-effects models} allowing for the incorporation of heterogeneous effects and behavioral differences accounting for the non-independent nature of longitudinal data at hand. 
Mixed-effects models include both \emph{fixed} and \emph{random} effects; following \cite{gelman2005analysis}, we refer to fixed effects as effects being constant across levels (e.g., individuals) and random effects as those varying between different levels. An overview of mixed-effects models can be found in \cite{pinheiro2006mixed}.

In our setting, the introduction of random effects enabled us to consider variations between different levels; the most important level being different users accounting for the inherent differences between individual Reddit users (e.g., the average quality of their comments).
As highlighted in \cite{baayen2008mixed}, mixed-effects models have further advantages, such as flexibility in handling (\emph{i})  missing data and (\emph{ii}) continuous and categorical responses, as well as (\emph{iii}) the capability of modeling heteroscedasticity.
 For simplicity, let us specify mixed-effects model equations using the following syntax \cite{bates}:
 
 \begin{equation}
 \text{outcome} \sim 1 + \text{fixed effect} + (\text{random effect}|\text{level})
 \label{eq:model}
 \end{equation}
 
 This specification describes a model where an outcome (dependent variable) is explained by an intercept $1$, one or more fixed effect(s), as well as one or more random effects allowing for variations between levels. 
 For all our experiments, we utilize the \emph{lme4} R package~\cite{bates} and fit the models with maximum likelihood. Examples about model specifications can be found online.\footnote{\url{http://glmm.wikidot.com/faq\#toc27}}

 \green{
As each of our experiments is conducted on one of our four different features that all exhibit different properties---e.g., count (text length) vs. continuous (readability) data---we performed extensive model analytics to find the most suitable model for each problem setting. 
Overall, we aimed at finding the most appropriate model for each feature at hand by not only focusing on simple linear mixed-effects models, but also on generalized mixed-effects models such as Poisson or negative Binomial regression suitable for count data. When fitting regression models, several assumptions need to be considered, such as for linear models we need to check for normally distributed residuals and heteroscedasticity. Thus, we performed model diagnostics on the individual models and successively tried to improve our models, for example going from a linear model to a Poisson model. Additionally, we checked for overdispersion and zero-inflation in our count data models (Poisson and negative binomial) and accounted for it. We also tackled problems like multicollinearity, outlier bias, as well as convergence problems. The models reported in this article are the ones that we judged as the most useful ones for each setting at hand after extensive model diagnostics outlined above.} 

 \green{
For judging significance of fixed and random effects, we followed an incremental modeling approach starting with the most simple model only explaining the outcome by the intercept and then subsequently adding effects to the model. For comparing the relative fits of these models we used the \emph{Bayesian Information Criterion (BIC)} \cite{schwarz1978estimating} which balances the likelihood of a model with its complexity. An interpretation table presented by Kass and Raftery \cite{kass1995bayes} can be consulted to determine the strength of the differences between BIC scores. This allows to gain confidence in the significance of observed effects allowing us to make inference on them. All reported fixed effects in this work are highly significant---except where mentioned (randomized baseline data)---meaning that the differences in BIC scores between the models including the effect and those excluding it are far larger than the maximum threshold of $10$ indicating strong evidence as postulated in \cite{kass1995bayes}. For completeness, we also conducted additional significance tests for the fixed effects such as t-tests or F-tests confirming our BIC diagnostics.
}

 \green{
In order to enable the reader to follow our individual steps and also allow for personal inference, we provide detailed reports for each experiment---based on a sample of $1$ million data points---in the form of jupyter notebooks using R kernels both online\fnref{notebooks} and in the supplementary material (S1-S8 Notebooks). 
In the main article, we only reported the fixed effects and corresponding inference as those were the main effects we were interested in. However, we make the full regression outputs available in the supplementary information (S1-S8 Tables).
Making our code and all experiments publicly available allows us to carefully document the results, as well as encourage other researchers to make their own inference and further refine our models.
At the same time, utilized Reddit data is freely available\fnref{data}.}

\bibliographystyle{abbrv}

\section*{Supplementary Material}

\paragraph*{S1 Notebook.}
\label{S1_Notebook}
{\bf Experimental steps for studying effects of session length on first comment's text length.} 

\paragraph*{S2 Notebook.}
\label{S2_Notebook}
{\bf Experimental steps for studying effects of session length on first comment's number of responses.} 

\paragraph*{S3 Notebook.}
\label{S3_Notebook}
{\bf Experimental steps for studying effects of session length on first comment's score.} 

\paragraph*{S4 Notebook.}
\label{S4_Notebook}
{\bf Experimental steps for studying effects of session length on first comment's readability.} 

\paragraph*{S5 Notebook.}
\label{S5_Notebook}
{\bf Experimental steps for studying effects of session index on a comment's text length.} 

\paragraph*{S6 Notebook.}
\label{S6_Notebook}
{\bf Experimental steps for studying effects of session index on a comment's number of responses.} 

\paragraph*{S7 Notebook.}
\label{S7_Notebook}
{\bf Experimental steps for studying effects of session index on a comment's score.} 

\paragraph*{S8 Notebook.}
\label{S8_Notebook}
{\bf Experimental steps for studying effects of session index on a comment's readability.} 

\paragraph*{S1 Table.}
\label{S1_Table}
{\bf Mixed-effects model results for effects of session length on first comment's text length.} This table presents the detailed mixed-effects model results for studying the effect of session length on the text length of the first comment $C_1$ in a session; i.e., data only contains the first session comments. The models at hand are linear mixed-effects models (lmer) where the outcome (text length) has been log-transformed. The baseline model excludes the fixed effect at interest for judging the significance of the effect; comparing the BIC of both models reveals a clear significance. This is confirmed by the AIC as well as the classic t-test on the coefficient.

\paragraph*{S2 Table.}
\label{S2_Table}
{\bf Mixed-effects model results for effects of session length on first comment's number of responses.} This table presents the detailed mixed-effects model results for studying the effect of session length on the number of responses of the first comment $C_1$ in a session; i.e., data only contains the first session comments. The models at hand are generalized linear  Poisson mixed-effects models (glmer) with a log link. The baseline model excludes the fixed effect at interest for judging the significance of the effect; comparing the BIC of both models reveals a clear significance. This is confirmed by the AIC as well as the classic t-test on the coefficient.

\paragraph*{S3 Table.}
\label{S3_Table}
{\bf Mixed-effects model results for effects of session length on first comment's score.} This table presents the detailed mixed-effects model results for studying the effect of session length on the score of the first comment $C_1$ in a session; i.e., data only contains the first session comments. The models at hand are generalized linear  Poisson mixed-effects models (glmer) with a log link---additionally we have added a constant for making the score always positive. The baseline model excludes the fixed effect at interest for judging the significance of the effect; comparing the BIC of both models reveals a clear significance. This is confirmed by the AIC as well as the classic t-test on the coefficient.

\paragraph*{S4 Table.}
\label{S4_Table}
{\bf Mixed-effects model results for effects of session length on first comment's readability.} This table presents the detailed mixed-effects model results for studying the effect of session length on the readability of the first comment $C_1$ in a session; i.e., data only contains the first session comments. The models at hand are linear mixed-effects models (lmer). The baseline model excludes the fixed effect at interest for judging the significance of the effect; comparing the BIC of both models reveals a clear significance. This is confirmed by the AIC as well as the classic t-test on the coefficient.

\paragraph*{S5 Table.}
\label{S5_Table}
{\bf Mixed-effects model results for effects of session index on a comment's text length.} This table presents the detailed mixed-effects model results for studying the effect of the session index $i$ on the text length of respective comment $C_i$; i.e., data includes all session comments. The models at hand are linear mixed-effects models (lmer) where the outcome (text length) has been log-transformed. The baseline model excludes the fixed effect at interest for judging the significance of the effect; comparing the BIC of both models reveals a clear significance. This is confirmed by the AIC as well as the classic t-test on the coefficient.

\paragraph*{S6 Table.}
\label{S6_Table}
{\bf Mixed-effects model results for effects of session index on a comment's number of responses.} This table presents the detailed mixed-effects model results for studying the effect of the session index $i$ on the number of responses of respective comment $C_i$; i.e., data includes all session comments. The models at hand are generalized linear  Poisson mixed-effects models (glmer) with a log link. The baseline model excludes the fixed effect at interest for judging the significance of the effect; comparing the BIC of both models reveals a clear significance. This is confirmed by the AIC as well as the classic t-test on the coefficient.

\paragraph*{S7 Table.}
\label{S7_Table}
{\bf Mixed-effects model results for effects of session index on a comment's score.} This table presents the detailed mixed-effects model results for studying the effect of the session index $i$ on the score of respective comment $C_i$; i.e., data includes all session comments. The models at hand are generalized linear  Poisson mixed-effects models (glmer) with a log link---additionally we have added a constant for making the score always positive. The baseline model excludes the fixed effect at interest for judging the significance of the effect; comparing the BIC of both models reveals a clear significance. This is confirmed by the AIC as well as the classic t-test on the coefficient.

\paragraph*{S8 Table.}
\label{S8_Table}
{\bf Mixed-effects model results for effects of session index on a comment's readability.} This table presents the detailed mixed-effects model results for studying the effect of the session index $i$ on the readability of respective comment $C_i$; i.e., data includes all session comments. The models at hand are linear mixed-effects models (lmer). The baseline model excludes the fixed effect at interest for judging the significance of the effect; comparing the BIC of both models reveals a clear significance. This is confirmed by the AIC as well as the classic t-test on the coefficient.

\end{document}